\documentclass[11pt]{article}
\usepackage{latexsym}
\usepackage{mathrsfs}
\usepackage{amssymb}
\usepackage{amsmath}
\usepackage{graphicx}
\usepackage{cite}

\usepackage[hidelinks]{hyperref}
\usepackage{cleveref}

\newcommand{\half}{\frac{\scriptstyle 1}{\scriptstyle 2}}
\newcommand{\A}{\mathbb{A}}
\newcommand{\C}{\mathbb{C}}

\newcommand{\CP}{\mathbb{CP}}

\newcommand{\F}{\mathscr{F}}

\newcommand{\T}{\mathbb{T}}

\newcommand{\p}{\partial}
\newcommand{\dbar}{\bar\partial}
\newcommand{\e}{\mathrm{e}}

\newcommand{\cZ}{\mathcal{Z}}

\newcommand{\rd}{\, \mathrm{d}}

\newcommand{\pf}{\mathrm{Pf}\,}
\newcommand{\be}{\begin{equation}\label}
\newcommand{\ee}{\end{equation}}
\newcommand{\bea}{\begin{eqnarray}\label}
\newcommand{\eea}{\end{eqnarray}}

\begin{document}

\title{The polarized scattering equations for 6d superamplitudes}
\author{Yvonne Geyer$^\dagger$, Lionel Mason$^\ddagger$ \vspace{.1cm}\\
%\address{
\small{$^\dagger$ Institute for Advanced Study, 1 Einstein Drive, 08540 Princeton NJ, USA}\\
\small{$^\ddagger$Mathematical Institute, University of Oxford, Woodstock Road, Oxford OX2 6GG, UK}}

\maketitle

\begin{abstract}
We introduce a spinorial version of the scattering equations,  the \emph{polarized scattering equations}, that incorporates spinor polarization data.  They lead to new formulae for tree-level scattering amplitudes in six dimensions that directly extend to maximal supersymmetry.  They give  a quite distinct framework from that of Cachazo et al. \cite{Cachazo:2018hqa}; in particular, the formulae do not change character from even to odd numbers of particles.  We find new ingredients for integrands for maximally supersymmetric  Yang-Mills, gravity, M5 and D5 branes. % together with rather speculative formulae for nonabelian theories with $(0,2)$ supersymmetry and gravitational theories with $(0,4)$ supersymmetry. 
 % except in the form of the integrands. 
We explain how the polarized scattering equations and supersymmetry representations arise from an ambitwistor-string with target given by a  super-twistor description of the geometry of super-ambitwistor space for six dimensions.  On reduction to four dimensions the polarized scattering equations give rise to massive analogues of the 4d refined scattering equations for amplitudes on the Coulomb branch.    At zero mass this framework naturally generalizes the twistorial version of the ambitwistor string in four dimensions. 

\end{abstract}

\section{Introduction}

%Although null geodesics are conformally invariant, this description is not.  A conformally invariant extension was provided in \cite{Adamo:2016} using the standard conformal compactification of $d$-dimensional Minkowski space as the quadric in in $d+1$-dimensional projective space.  
%Introduce $d+2$-homogeneous coordinates $X^M=(x^+,x^-,x^\mu)$ on $\CP^{d+1}$, then the compactification is the quadric $\{X^MX_M=x^+x^-+x^\mu x_\mu=0\}\subset \CP^{d+1}$.

Six dimensions has proved to be a fertile arena both for unifying four dimensional field theories via dimensional reduction and studying lower dimensional consequences of M-theory.  In the context of scattering amplitudes, a first exploration of frameworks inspired by the twistor-string and the scattering equations appeared in \cite{Heydeman:2017yww,Cachazo:2018hqa}.  These works gave an extension of the 4d formulae of \cite{Roiban:2004yf, Cachazo:2012kg} arising from
twistor string theories \cite{Witten:2003nn,Berkovits:2004hg,Skinner:2013xp} to 6d for a variety of theories.  Formulae for the bosonic parts of these theories  had already been found in all dimensions by Cachazo, He and Yuan (CHY) \cite{Cachazo:2013hca, Cachazo:2014dia} including D-branes and Born-Infeld theories, but without manifest supersymmetry or fermion external states. These CHY formulae were seen to arise from chiral string theories in the space of complex null geodesics, ambitwistor space \cite{Mason:2013sva}.  This theory leads to a procedure for constructing fermionic amplitudes using its Ramond sector \cite{Adamo:2013tsa} as  does the supersymmetric pure spinor formulation in 10 dimensions \cite{Berkovits:2013xba}, but neither lead to such explicit formulae for arbitrary $n$-point superamplitudes, partly because supermomenta  cannot be defined in a globally Lorentz invariant manner in 10d.  

A different formulation was obtained for 4d superamplitude formulae in \cite{Geyer:2014fka} by applying the ambitwistor perspective to a twistorial coordinatization of ambitwistor space in 4d.  This led to optimally compact and explicit expressions for superamplitudes. These formulae were  based on an extension of the scattering equations that are not only refined by MHV degree, but also incorporate the polarization data. It is this approach that we extend  to six dimensions.  Our formulae are motivated by a worldsheet model with target space a chiral twistorial representation of 6d ambitwistor space.  Generic path integrals involving momentum eigenstates in this model are supported on \emph{polarized scattering equations} -- constraints formulated in terms of worldsheet spinors with little-group indices and depending on the polarization data of the particles.  The model has a natural supersymmetric extension  that leads to compact expressions for superamplitudes.  However, it is awkward to express the model in a self-contained way for non-chiral theories so details are only sketched briefly here as motivation for our formulae.

  The formulae we obtain sidestep an awkwardness of those in \cite{Cachazo:2018hqa} that require a very different treatment for amplitudes with odd numbers of particles to those for even numbers.  We furthermore introduce new integrands that provide the natural 6d extension of the Hodges determinants arising in 4d, rather than the CHY Pfaffians (although they still play a role in our D5- and M5-brane formulae).   Many of the  ingredients in our formulae have a clear worldsheet origin. Our supersymmetric formulae do not break R-symmetry following the observation that the polarized scattering equations give a natural breaking of little group symmetry (this is related to proposals in  
\cite{Huang:2010rn, Bandos:2014lja}).  
% Huang:2010rn,

We start with a presentation of 6d spinor and polarization data in \S\ref{sec:polda},  the polarized scattering equations in \S\ref{sec:polSE}, together with a chiral measure for the formulae.  In \S\ref{sec:susy} we introduce the factors encoding the supermomenta and in \S\ref{sec:integrands} the remaining integrands that make up the formulae. In \S\ref{sec:ambitwistor-string} we explain the origin of the various ingredients in a worldsheet model with target a twistorial representation of ambitwistor space. As an application in 4d, in \S\ref{sec:CB} we explain how the formulae reduce to give formulae for massive amplitudes in 4d.  Finally in \S\ref{sec:discussion} we discuss further checks by reduction to 4d, give a table for the various formulae we can obtain, including the rather speculative ones for nonabelian $(2,0)$ and gravitational $(4,0)$ theories. We plan to publish full details of many of these ideas in a sequel.

\section{Spinors and polarization data in 6 dimensions}\label{sec:polda}
In six dimensions, the spin group in the complex is $\mathrm{SL}(4,\C)$  and  acts on vectors as skew spinors $k^{AB}=k^{[AB]}$
%=\gamma_\mu^{AB}k^\mu$
  where $A, B=0,\ldots , 3$ are spinor indices. The metric is determined by the totally skew spinor $\varepsilon_{ABCD}=\varepsilon_{[ABCD]}$ which is also used to raise and lower skew pairs of spinor indices.  Since the skew spinors are six dimensional, we can do away with Clifford matrices and work with skew pairs of spinor indices instead of vector indices throughout.
% and $\gamma_\mu^{AB}$ are antisymmetric $4\times4$ matrices satisfying the Clifford algebra.  The inner product of vectors is defined via the totally skew, $\mathrm{SU}(4)$-invariant tensor $\varepsilon_{ABCD}$,
 
For massless particles, the little group is  $\mathrm{Spin}(4)_\mathbb{C}\cong\mathrm{SL}(2)_+\times\mathrm{SL}(2)_-$ which can be seen as follows. 
Null momenta $k^{AB}$ with $k^2=k^{AB}k^{CD}\varepsilon_{ABCD}=0$ are necessarily of rank two due to the antisymmetry of the spinor indices.  They can therefore be represented either by chiral or by antichiral spinors \cite{Cheung:2009dc} as
\begin{equation}
k^{AB}= \varepsilon ^{\dot a\dot b} \kappa_{\dot a}^A\kappa_{\dot b}^B=:\left[\kappa^A\kappa^B\right]\,,\qquad k_{AB}= \kappa^{ a}_C \kappa^{b}_D \varepsilon_{ a b}=:\left\langle\kappa_A\kappa_B\right\rangle\,.\label{eq:momenta}
\end{equation}
Here, $a=0,1$, $\dot a=\dot 0, \dot 1$ are the corresponding $\mathrm{SU}(2)_{\pm}$ little group spinor indices, and we have introduced the four-dimensional $\langle\cdot,\cdot \rangle$ and $[\cdot,\cdot]$ brackets for little group contractions.  %The spinor-helicity decomposition \eqref{eq:momenta} can also be recast in the useful relation
%\begin{equation}
%\kappa_{[A}^a\kappa_{B]}^b=\varepsilon^{ab} k_{AB}\,,\qquad \kappa^{[A}_{\dot a}\kappa^{B]}_{\dot b}=\varepsilon_{\dot a\dot b} k^{AB}\,.\label{useful}
%\end{equation}

The spinor-helicity formalism represents the polarization data as follows. A Maxwell field strength is represented by $F^A_B$, with $F_A^A=0$ because the Lie algebra  of the Lorentz group is $\mathrm{sl}(4)$. We take momentum eigenstates with null and transverse polarization vectors giving
\begin{equation}
F^A_B=\epsilon^A\epsilon_B\,.\label{eq:F}
\end{equation}
The Maxwell equations yield $k_{AB}\epsilon^A=0=k^{AB}\epsilon_B$, so that the polarization data is encoded in little group spinors $\epsilon_a$ and $\epsilon_{\dot a}$ with
%\footnote{Note that $\epsilon_a$ and $\epsilon_{\dot a}$ cannot be taken to be real in Lorentz signature.}
\begin{equation}\label{eq:pol}
\epsilon^A=\epsilon_{\dot a}\kappa^{A\dot a}\, , \qquad \epsilon_A=\epsilon^a\kappa_{Aa}\, .
\end{equation}

\section{The polarized scattering equations}\label{sec:polSE}
Consider $n$ external null momenta $k_i$, $i=1,\ldots , n$ with $\sum_i k_i=0$ and associated polarization spinors $\epsilon_{iA}, \epsilon_i^A$.
In the CHY formulae and the ambitwistor string, a key role is played by the meromorphic vector on the Riemann sphere with complex coordinate $\sigma\in \C$
\begin{equation}
P(\sigma)=\sum_i \frac{k_i}{\sigma-\sigma_i} \label{P-def}
\end{equation}
where $\sigma_i$ are $n$ marked points on $\CP^1$. Amplitudes in massless theories are then written as moduli space integrals, localized on so-called scattering equations.
The scattering equations are the condition that $P(\sigma)$ is a null vector for all $\sigma$ and are equivalent to the $n$ equations on the $\sigma_i$ given by 
\begin{equation}
k_i\cdot P(\sigma_i)=0\, .\label{SE}
\end{equation}
For such a null $P$, we can seek spinor-helicity factorizations  $\lambda^{a}_A(\sigma)$ and $\lambda^A_{\dot a}(\sigma)$ of $P(\sigma)$ 
\begin{equation}\label{P-fact}
P_{AB}=\lambda_{aA}\lambda_{B}^a=\half\varepsilon_{ABCD}\lambda^C_{\dot a}\lambda^{D\dot a}\, 
\end{equation}  
on $\CP^1$.
Motivated by the ambitwistor string model in \S\ref{sec:ambitwistor-string}, we seek $\lambda_{Aa}(\sigma)$ of the form
\begin{equation}
\lambda_{aA}(\sigma)=\sum_{i=1}^n\frac{u_{ia}\epsilon_{iA}}{\sigma-\sigma_i}\, .\label{lambda-def}
\end{equation}
The scattering equation yield $k_i\cdot P=\mathrm{det} (\kappa^a_{iA},\lambda^b_A)=0$ as a condition on their constituent spinors.   This determinant  vanishes iff there exists non zero $(u^a_i,v^a_i)$ defined up to scale so that 
\begin{equation}
u_{ia}\lambda^a_A(\sigma_i) =v_{ia}\kappa_{iA}^a\, .
 \label{6dSE+}
\end{equation}
This is scale invariant in $u$ and $v$, so we introduce the normalization $\langle v_i\epsilon_i\rangle=1$. Using this normalization,  \eqref{6dSE+} gives a deterministic system of equations in the variables $(\sigma_i, u_{ia}, v_{ia})$, and provides our six-dimensional \emph{polarized scattering equations}
\begin{equation}
\sum_{j=1}^n \frac{\langle u_i u_j\rangle \epsilon_{jA}}{\sigma_{ij}}-\langle v_i \,\kappa_{iA} \rangle=0\, .\label{polscatt}
\end{equation}
These equations are non-singular  because $\varepsilon^{ab}$ is skew.  There exists a unique solution for each solution $\{\sigma_i\}$ to the unpolarized scattering equations \eqref{SE}. Unlike \cite{Heydeman:2017yww, Cachazo:2018hqa},  who introduce different constructions for even and odd numbers of particles, our polarized scattering equations do not distinguish between odd and even numbers of particles.

It is easy to see that \eqref{P-fact} holds:
  there are no double poles as $u^a_iu_{ia}=0$ and 
$\mathrm{Res}_{\sigma_i} \lambda(\sigma)_{A}^a\lambda_{Ba}= \epsilon_{i[A}u_{ia} \lambda_{B]}^a(\sigma_i)=v_i^b\epsilon_{ia}\kappa^a_{[A} \kappa_{iB]b}=:k_{iAB}$ on the support of the polarized scattering equations and the normalization condition $\langle \epsilon v\rangle=1$. 

\smallskip

The polarized scattering equations enhance the $\mathrm{SL}(2,\mathbb{C})_\sigma$ symmetry of the unpolarized scattering equations to a global $\mathrm{SL}(2,\mathbb{C})_\sigma \times \mathrm{SL}(2,\mathbb{C})_+$ symmetry, corresponding to M\"obius invariance and (complexified, chiral) little group invariance respectively. Taking this into account, we can define a covariant measure $\rd\mu_n^{\mathrm{pol}}$ via
\begin{equation}\label{eq:measure}
 \rd\mu_n^{\mathrm{pol}}\equiv \frac{\prod_{i=1}^n \rd \sigma_i\,\rd^2 u_i\,\rd^2 v_i}{\mathrm{vol}\; \mathrm{SL}(2,\mathbb{C})_\sigma \times \mathrm{SL}(2,\mathbb{C})_+}\prod_{i=1}^n \delta\big(\langle v_i \varepsilon_i \rangle -1 \big)\;\delta^4\Big(\langle u_{i}\lambda_A(\sigma_i)\rangle -\langle v_{i}\kappa_{iA}\rangle\Big)\,.
\end{equation}
This measure contains $5n$ delta-functions on the $5n-(3+3)$ variables $(u_i,v_i,\sigma_i)$ modulo the symmetries,  leaving an overall 6d momentum-conserving delta-function. The polarized scattering equations imply the scattering equations \eqref{SE}, and the $(u_i,v_i)$ can be explicitly integrated out to yield  $\rd\mu_n^{\mathrm{pol}}=\delta^6(\sum_i k_i)\rd\mu_n^{\mathrm{CHY}}$ where $\rd\mu_n^{\mathrm{CHY}}$ is the standard CHY measure $\prod_i \delta(k_i\cdot P(\sigma_i))\rd\sigma_i/\mathrm{SL}(2,\mathbb{C})$. For ambidextrous theories, the conjugate scattering equations $[\tilde  u_{i}\lambda^A(\sigma_i)] -[\tilde  v_{i}\kappa_{i}^A]=0$ are imposed, but not included in the measure, in line with \cite{Cachazo:2018hqa, Heydeman:2017yww}. 

We will express six-dimensional scattering amplitudes  as, for example
\begin{equation}\label{eq:form_amplitude}
 \int \rd\mu_n^{\mathrm{pol}}\,\mathcal{I}_n\,,\qquad \text{with }\;\mathcal{I}_n=\begin{cases}I_n^{\mathrm{spin-1}}\; \mathrm{PT}(\alpha) & \;\text{Yang-Mills}\\I_n^{\mathrm{spin-1}}\;\tilde{I}_n^{\mathrm{spin-1}} & \;\text{gravity}\,,\end{cases}
\end{equation}
where the integrand $\mathcal{I}_n$ is a holomorphic function of the scattering data and the variables $(\sigma_i, u_{ia}, v_{ia})$ carrying $\mathrm{SL}(2,\mathbb{C})_\sigma $ weight $2$. Here $\mathrm{PT}(\alpha)=\prod(\sigma_{\alpha(i)}-\sigma_{\alpha(i+1)})^{-1}$ is the Parke-Taylor factor associated to the permutation $\alpha$ and  $I_n^{\mathrm{spin-1}}$ is a function of Maxwell polarization data that is gauge and permutation invariant on each solution to the polarized scattering equations. If we were not interested in supersymmetry, we could insert for example the CHY Pfaffians for $I_n^{\mathrm{spin-1}}$ but this of course does not exploit the new ingredients arising from the polarized scattering equations.  These can be introduced as follows.

\section{Supersymmetry}\label{sec:susy}
In six dimensions, $(N,\tilde N)$-supersymmetry  possesses an $\mathrm{Sp}(N)\times\mathrm{Sp}(\tilde N)$ R-symmetry group for which we introduce indices $I=1,\ldots , 2N,$ and $\dot I=\dot 1, \ldots,\dot {2\tilde N}$.   
On momentum eigenstates with momentum $k_{AB}$, the supersymmetry generators $Q_{AI}$ and $Q^A_{\dot I}$ satisfy 
\begin{equation}
\{Q_{A I},Q_{B J}\}=k_{AB} \, \Omega_{IJ},\qquad \{Q^A_{\dot I},Q^B_{\dot J}\}=k^{AB} \, \Omega_{\dot I\dot J}
\end{equation} where $\Omega_{IJ}$ and $\Omega_{\dot I \dot J}$ are the R-symmetry symplectic metrics.
The supersymmetry generators thus reduce to the little group as 
\begin{equation}
 Q_{A I}=\kappa_A^a Q_{aI}\,,\qquad Q^{A}_{\dot I}=\kappa^A_{\dot a} Q_{\dot I}^{\dot a}
\end{equation}

A key example is maximal super Yang-Mills theory.  This has $(1,1)$-supersymmetry with supermultiplet given by\footnote{In the following, we suppress the particle index $i$ for readability.}
\begin{equation}
\F:=(F_A^B, \,\psi^A_{I},\,\tilde\psi_{A \dot I},\, \phi_{I\dot I})\, .
\end{equation}
On momentum eigenstates with null momentum $k_{AB}$,  $Q_{C J}$ and $Q^C_{\dot J}$ act  by
\begin{eqnarray}
Q_{C J}\F&=&(k_{AC}\psi^B_{J},\, \Omega_{JI}F_C^A, \,k_{AC}\phi_{J\dot I}, \,\Omega_{JI} \tilde \psi_{\dot I C})\, , \nonumber\\
Q^C_{ \dot J}\;\F&=&(k^{BC}\tilde \psi_{A\dot J}, \,k^{BC}\phi_{\dot J I},\,\Omega_{\dot J\dot I} F_A^C, \, \Omega_{\dot J\dot I}  \psi^C_{ I })\, .
\end{eqnarray}

To construct an on-shell superspace, half of the generators need to be selected as supermomenta. Two mechanisms have been discussed in the literature \cite{Huang:2010rn}, manifesting either little-group or R-symmetry. While the former has been employed successfully in recent work on 6d scattering amplitudes in a variety of theories \cite{Cachazo:2018hqa, Heydeman:2017yww}, the latter is more natural from the perspective of the ambitwistor string \cite{Bandos:2014lja}, and will be the formulation we work with here. The two approaches are of course related by appropriate Grassmann Fourier transforms.

\smallskip

In the context of the polarized scattering equations, $\epsilon_a$, $\epsilon_{\dot a}$ and $v_{a}$, $v_{\dot a}$ give a natural choice for breaking the little-group symmetry.  Using this, the full supermultiplet can be parametrized by supermomenta $(q_I,\tilde q_{\dot I})$ by imposing the relations
\begin{equation}\label{eq:def_susy-gen}
Q_{aI}\F(q,\tilde q)=\left(v_a q_I +\epsilon_a\,\Omega_{IJ}\frac{\p}{\p q_J}\right) \F(q,\tilde q)\, .
\end{equation}
In this representation, the full super Yang-Mills  multiplet can be obtained from the pure gluon state $\F(0,0)=(\epsilon_A\epsilon^B,0,0,0)$. In particular, the multiplet is parametrized by
\begin{equation}
\F(q_I,\tilde q_{\dot I})=\left((\epsilon_A+ q^2\langle v \kappa_A\rangle)(\epsilon^B+ \tilde q^2 \langle v \kappa^B\rangle), \,q_I(\epsilon^A+\tilde q^2 \langle v \kappa^A\rangle),\, \tilde q_{\dot I}(\epsilon_A+q^2 \langle v \kappa_A\rangle),\, q_I\tilde q_{\dot I}\right)\, .
\end{equation}
While this supersymmetry representation is dynamic and particle-specific, the full supersymmetry generator for $n$ particles is still  defined by the sum  $Q_{AI}=\sum_{i=1}^n Q_i{}_{AI}$.  Superamplitudes --  as supersymetrically invariant objects -- are annihilated by the supersymmetry generators $Q_{AI}$. As we will verify below, this implies that the total dependence on the supermomenta is encoded in a simple exponential  factor $\e^{F}$, with $F=F_{\scalebox{0.7}{$N$}}+\tilde{F}_{\scalebox{0.7}{$\tilde N$}}$ and
\begin{equation}
 F_{\scalebox{0.7}{$N$}}=\sum_{i<j}   \frac{ \langle u_i u_j\rangle q_{iI}q_j^I}{\sigma_{ij}}\,,\qquad \tilde{F}_{\scalebox{0.7}{$\tilde N$}}=\sum_{i<j}   \frac{ [ \tilde u_i \tilde u_j] \tilde{q}_{i\dot I}\tilde{q}_j^{\dot I}}{\sigma_{ij}}\,.
\end{equation}
For $\mathcal{N}=(1,1)$ super Yang-Mills for example, this exponential factor becomes $\exp F^{\mathrm{YM}}=\exp (F_1+\tilde{F}_1)$. We will also provide a geometric derivation of the factor $\e^{F_{\scalebox{0.6}{$N$}}}$ from a worldsheet model in \cref{sec:ambitwistor-string}.

In general,  given a scattering amplitude of the form \eqref{eq:form_amplitude} for the top states of the multiplet of an $\mathcal{N}=(N,\tilde N)$ theory, the fully supersymmetric amplitude is given by
\begin{equation}\label{eq:ampl_susy}
 \mathcal{A}_n=\int \rd\mu_n^{\mathrm{pol}}\,\mathcal{I}_n\;e^{F}\,.
\end{equation}
It is easily verified that this amplitude  is supersymetrically invariant, since
\begin{equation}
 Q_{AI} \,e^{F}=\left(\sum_i \langle v_i\kappa_{iA}\rangle \,q_{iI}-\sum_{i,j}\frac{\langle u_i u_j\rangle\,\epsilon_{iA}}{\sigma_{ij}}q_{jI}\right)e^{F}=0\,,
\end{equation}
on the polarized scattering equations, and similarly $Q^A_{\dot I} \,e^{F}=0$.  Reversely, given an integrand $\mathcal{I}_n$ for the top states of a multiplet,  \eqref{eq:ampl_susy} is the unique supersymmetric completion using the supersymmetry representation \eqref{eq:def_susy-gen}, as can be verified using supersymmetric Ward identities.

\section{Integrands}\label{sec:integrands}
We now construct the integrands $\mathcal{I}_n$ for SYM, supergravity, D5 theory and M5 theory. 
For the ambidextrous spin one contribution, define an $n\times n$  matrix $H$ by 
\begin{equation}
H_{ij} =\begin{cases}\frac{\epsilon_{iA}\epsilon^A_j}{\sigma_{ij}} \qquad i\neq j\\
\label{eq:def1_H_ii}
 H_{ii}:=-e_i\cdot P(\sigma_i)\, ,\end{cases}
\end{equation}
where $e_i=\epsilon^{[A}_i\hat \epsilon^{B]}_i$  is the polarization vector and $\hat \epsilon_i^{B}k_{iAB}=\epsilon_{iA}$. We have
\begin{equation}\label{eq:def2_H_ii}
\lambda_{aA}(\sigma_i)\epsilon^A_i=-u_{ia} H_{ii} \, , \quad \mbox{ or  }\quad \lambda^{\dot aA}(\sigma_i) \epsilon_{iA}=-u_i^{\dot a} H_{ii} \,.
\end{equation}
The fact that the LHS is a multiple of $u_{ia}$ follows from the scattering equation $u_{ia}\lambda^a_A(\sigma_i)=v^a\kappa_{aA}$ and the identity $k^{AB} \kappa_{aA}=0$. 
Starting from the \eqref{eq:def2_H_ii} we recover \eqref{eq:def1_H_ii} from
\begin{multline}
e_i\cdot P(\sigma_i)= \epsilon_i^{[A}\hat \epsilon_i^{B]}\lambda_{aA}(\sigma_i)\lambda^a_B(\sigma_i)=-H_{ii}\, \hat \epsilon_i^{B}u_{ia}\lambda^a_B(\sigma_i)=-H_{ii} \hat \epsilon^{B}_iv_{ia}\kappa^a_{iB}=-H_{ii}\, . 
\end{multline}

On the polarized scattering equations, the determinant $\det H$ vanishes because the matrix $H_{ij}$ is not full rank due to
\begin{equation}
\sum_i u_{ia} H_{ij}=\lambda_{a A} (\sigma_j)\epsilon^A_j +u_{ja} H_{jj}=0\, .
\end{equation}
The first term follows from the definition \eqref{lambda-def} of $\lambda_{a A} $ and the second equality from \eqref{eq:def2_H_ii}. Similarly,  $\sum_j H_{ij} u_{j\dot a}=0$. These identities nevertheless imply that $H$ has a well defined  reduced determinant 
\begin{align}
\det{}'H: =\frac{\det(H^{[i_1i_2]}_{[j_1j_2]})}{\langle u_{i_1} u_{i_2}\rangle [u_{j_1} u_{j_2}]}\,.\label{gendet2}
\end{align}
Here $H^{[i_1j_1]}_{[i_2j_3]}$ denotes the matrix $H$ with the rows $i_1,\,i_2$ and columns $j_1,\,j_2$ deleted, and $\det{}'H$ is well-defined in the sense that the \eqref{gendet2} is invariant under permutations of particle labels, and thus independent of the choice of $i_{1,2},\,j_{1,2}$. 

To see this, we extend the argument of appendix A of \cite{Cachazo:2012pz}  to reduced determinants of general matrices.  Since the  $n\times n$ matrix $H_{i}^{j}$ satisfies $\sum_i u^{i}_aH_{i}^{j}=0$ and $\sum_j H_{i}^{j}\tilde u_{j}^{\dot b}=0$, we can write
 \begin{equation}
 \varepsilon^{i_{1}\ldots i_n}\varepsilon_{j_1\ldots j_n}H_{i_{3}}^{j_{3}}\ldots H_{i_{n}}^{j_{n}}=\det{}'(H) \varepsilon^{a_1 a_2} u^{[i_{1}}_{a_1} u_{a_2}^{i_2]} \varepsilon_{\dot b_1\dot b_2}\tilde{u}^{\dot b_1}_{[j_1}\tilde{u}^{\dot b_2}_{j_2]} \label{red-det}
 \end{equation}
 for some $\det' H$ because the $u$'s and $\tilde u$'s span the left and right kernel of $H$.
 The definition \eqref{gendet2} then follows by taking components  of this definition on the $i_{1}, i_2, j_{1}, j_2$ indices.

The reduced determinant $\det{}'H$ is manifestly gauge invariant in all particles, carries $\mathrm{SL}(2,\mathbb{C})_\sigma$ weight $-2$, as expected for a half-integrand $I^{\mathrm{spin-1}}$ and is equally valid for even and odd numbers of external particles. On the support of the polarized scattering equations, it can be verified using factorization that $\det{}'H$ is equal to the CHY half-integrand $\pf{}'M$.
\smallskip

Another important building block, relevant for the D5 and M5 theory, is the skew matrix $A$, familiar from the CHY formulae \cite{Cachazo:2013hca, Cachazo:2014xea}, with
\begin{equation}
 A_{ij}=\frac{k_i\cdot k_j}{\sigma_{ij}}\,.
\end{equation}
Again, the Pfaffian $\mathrm{Pf} A$ vanishes as $A$ is degenerate due to the scattering equations \eqref{SE}, but the reduced Pfaffian $\mathrm{Pf}'A=\frac{(-1)^{i+j}}{\sigma_{ij}}\mathrm{Pf} A^{ij}_{ij}$ is well-defined and non-zero for even numbers of particles  \cite{ Cachazo:2013hca,Cachazo:2014xea}.

\smallskip

The final two ingredients are constructed exclusively using the variables $(\sigma_i, u_{ia}, \tilde u_{i\dot a})$, and are only needed for  M5-branes.  These only lead to amplitudes with even  numbers of particles. Consider any partition of the particle labels into two sets\footnote{For scalar amplitudes, this corresponds to a partition into two non-self-interacting scalar sectors $Y$ and $\overline{Y}$ with $n/2$ scalars of type $\phi_{1\dot 1}$ and $n/2$ of type $\phi_{2\dot 2}$.} $Y$ and $\overline{Y}$, with $|Y|=|\overline{Y}|=n/2$, and define $n/2\times n/2$ matrices $U$, $\tilde U$ and $X$ by
\begin{equation}\label{eq:def_Ured}
 U_{ir}=\frac{\langle u_iu_r\rangle}{\sigma_{ir}}\,,\qquad \tilde{U}_{ir}=\frac{[\tilde{u}_i\tilde{u}_r]}{\sigma_{ir}}\,,\qquad X_{ir}=\frac{1}{\sigma_{ir}}\,,
\end{equation}
for $i\in Y$ and $r\in\overline{Y}$. Schouten identities then guarantee that the ratios 
\begin{equation}
 \frac{\det \tilde U}{\det U}\,,  \qquad \frac{\det X}{\det{}^2 U}\,,\qquad \frac{\det X}{\det U\, \det \tilde U}\,,
\end{equation}
are independent of the choice of partition into $Y$ and $\overline{Y}$. Then the combinations
\begin{equation}
 \frac{\det X}{\det{}^2 U}\;\pf{}'A\,,\qquad \frac{\det X}{\det U\,\det \tilde U}\;\pf{}'A=\det{}'H\,,
\end{equation}
form chiral half-integrands, independent of the partition $Y, \overline{Y}$.  The latter equality follows from comparison to the CHY integrands when restricted to a scalar subsector.

\smallskip

At this point, we have all the ingredients to present the integrands of $\mathcal{N}=(1,1)$ super Yang-Mills, $\mathcal{N}=(2,2)$ supergravity, $\mathcal{N}=(1,1)$ D5 theory and $\mathcal{N}=(2,0)$ M5 theory;
\begin{subequations}\label{eq:integrands}
 \begin{align}
  & \text{Super Yang-Mills:} && \mathrm{PT}(\alpha)\;\det{}'H\;e^{F_1+\tilde{F}_1} \label{eq:int_sYM}\\
  & \text{Supergravity:} && \det{}'H\;\det{}'\tilde H\;e^{F_2+\tilde{F}_2}\\
  & \text{D5-branes:} && \det{}'A\;\det{}'H\;e^{F_1+\tilde{F}_1} \\
  & \text{M5-branes:} && \frac{\det X}{\det{}^2 U}\left(\pf{}'A\right)^3\;e^{F_2}
 \end{align}
\end{subequations}
It is easily seen that the resulting superamplitudes are $\mathrm{SL}(2,\mathbb{C})_\sigma\times \mathrm{SL}(2,\mathbb{C})_\pm$ invariant. As discussed above, the SYM and supergravity amplitudes are gauge invariant, and the supergravity amplitudes are permutation invariant. Moreover, the super Yang-Mills and supergravity amplitudes satisfy colour-kinematics duality, as evident from the form of their amplitudes, and the M5 amplitudes are manifestly chiral. We have checked that all amplitudes factorize correctly, and that the supergravity and super Yang-Mills amplitudes reduce to the well-known amplitudes  \cite{Witten:2003nn,Berkovits:2004hg,Roiban:2004yf, Roiban:2004ka} in four dimensions.

\section{Ambitwistor Strings}\label{sec:ambitwistor-string}
In the following, we develop a twistorial 6d ambitwistor string -- a chiral 2d CFT whose  target space is ambitwistor space $\A$, the space of complex light rays in complexified 6d space-time. This provides a geometric origin of the polarized scattering equations and the supersymmetry representation introduced in sections \ref{sec:polSE} and \ref{sec:susy}.

We focus here on the model using a chiral representation of (super) ambitwistor space. While this gives a derivation of the measure $\rd\mu^{\mathrm{pol}}_n$ and the chiral supersymmetry factors $\e^{F_{\scalebox{0.6}{$N$}}}$, amplitudes in ambidextrous theories with $(N,\tilde N)$-supersymmetry 
 require both the chiral supertwistors and their antichiral versions.  However, since these are alternative coordinates on the same space, they are related by non-trivial constraints. 
Implementing these constraints at the level of the worldsheet model, as well as  adding `worldsheet matter' systems giving rise to the integrands $\mathcal{I}_n$, is beyond the scope of this letter.

\smallskip

In $d$ dimensions, ambitwistor space has a representation as  a Hamiltonian quotient of the cotangent bundle  \cite{Mason:2013sva}.  With coordinates $(P_\mu,x^\mu)$, %$\mu=0,\ldots , d-1$ in $d$-dimensions,
$P_\mu$ is the null momentum vector of the null geodesic through $x^\mu$, and $\A$ is  the Hamiltonian quotient of $T^*M$ by $P^2$,
\begin{equation}\label{A-defn}
\A=T^*M_{P^2=0}/\{ P^\mu \p/\p x^\mu\}\,.
\end{equation}
As in four dimensions \cite{Geyer:2014fka},  in six dimensions $\A$ has a canonical twistor representation as follows.  Projective twistor space $\T$ is the quadric in $\CP^7$ with homogeneous coordinates   \cite{Penrose:1986ca}
\begin{equation}
Z=(\lambda_A, \mu^A)\in \mathbb{S}_A\oplus \mathbb{S}^A \quad \mbox{ subject to }\quad Z\cdot Z:= \lambda_A\mu^A=0.
\end{equation} 
A twistor $Z$ determines  a totally null self-dual 3-plane in space-time via the incidence relations
$
\mu^{A} =x^{AB}\lambda_B\, .\label{6dinc}
$
Generically such three-planes are disjoint, but they do intersect along a null geodesic when the corresponding twistors satisfy $Z^0\cdot Z^1=0$.  
Projective ambitwistor space $\mathbb{PA}$, the space of unscaled null geodesics, can thus be represented by a pair of twistors i.e., $Z^a=(\lambda_A^a,\mu^{Aa})$ subject to $Z^a\cdot Z^b=0$ and modulo the symmetry group $\mathrm{SL}(2,\C)_+$ acting on the $a$ index.  The original definition \eqref{A-defn} can be recovered via  the incidence relation
\begin{equation}
\label{A6dinc}
\mu^{Aa} =x^{AB}\lambda_B^a\,, \qquad\text{ and }\;\;P_{AB}=\langle\lambda_A\lambda_B\rangle\,.
\end{equation}
%This follows as, for fixed $Z^a$, the condition $Z^0\cdot Z^1=0$ guarantees a non-trivial intersection of the corresponding 3-planes, which is then along a null geodesic.  
The constraint $Z^a\cdot Z^b=0$ implies the existence of $x^{AB}$ satisfying \eqref{A6dinc}, which is easily seen to be unique up to the addition of multiples of $P^{AB}$. %Moreover, the resulting null geodesic is unique up to $\mathrm{SL}(2,\C)_+$.

\subsection{Ambitwistor strings and polarized scattering equations}

The bosonic ambitwistor action of \cite{Mason:2013sva} is
\begin{equation}
S=\int_\Sigma P\cdot \dbar X + \frac{\tilde e}{2} P^2\,.
\end{equation}
Here, $X:\Sigma\rightarrow M$ is an embedding into complexified space-time, $P_\mu$ a holomorphic 1-form on $\Sigma $ with values in the pullback of $T^*M$, and $\tilde e\in\Omega^{(0,1)}\otimes T\Sigma$ acts as a Lagrange multiplier for the constraint $P^2=0$.  This constraint is automatically solved by $P_{AB}=\langle\lambda_A\lambda_B\rangle$, and the incidence relation \eqref{A6dinc} further gives  $P\cdot \bar\p X=\epsilon_{ab}\,Z^a\cdot \dbar Z^b$, where $\bar\p$ must include the gauge field $A_{ab}$ on quantities with little group indices $a$.  Thus we can take the twistors $Z^a$  to be worldsheet-spinors, with weight $-1$ in homogeneous worldsheet coordinates $\sigma$ and we obtain the action\footnote{An alternative conformally invariant ambitwistor action critical in 6 dimensions was given in \cite{Adamo:2017zkm}.}
\begin{equation}\label{eq:action}
S_{\scalebox{0.7}{$\mathbb{A}$}}=\int_\Sigma \epsilon_{ab}\,Z^a\cdot \dbar Z^b+ A_{ab\;}Z^a\cdot Z^b\,.
%=\int_\Sigma \epsilon_{ab}(\lambda^a_A \bar\p\mu^{bA}  -\mu^{aA}\bar\p\lambda^b_A) + A_{ab}\;Z^a\cdot Z^b\, .
\end{equation}
 In this action, we have already gauge-fixed worldsheet gravity, leading to the usual $\mathrm{SL}(2,\mathbb{C})_\sigma$ quotient in the measure. The  field $A_{ab}=A_{[ab]}\in\Omega^{(0,1)}$ gauges the little group $\mathrm{SL}(2,\C)_+$ symmetry, reducing the target space from  two copies of twistor space to ambitwistor space.\footnote{The $(\lambda,\mu)$ are $2\times 2\times 4=16$ variables subject to 3 constraints and a quotient by $\mathrm{SL}(2,\C)_+$ reducing to 10 dimensions, the correct answer for nonprojective ambitwistor space} The bosonic ambitwistor  string is a chiral CFT that becomes free and linear after BRST gauge-fixing $A_{ab}=0$. 

%[For the ambidextrous Yang-Mills theory and gravity models discussed below, we also need to introduce conjugate twistors $Z^{\dot a}=(\lambda^{A\dot a},\mu_A^{\dot a})$ with $S=\int_\Sigma \epsilon_{\dot a \dot b}\,Z^{\dot a}\dbar Z^{\dot b} + A_{\dot a\dot b}\;Z^{\dot a}\cdot Z^{\dot b}$.]

\smallskip

Vertex operators in the ambitwistor string are built from ambitwistor representatives for space-time fields, which are Dolbeault cohomology classes  in $H^1(P\A)$.   The integrated vertex operators are given by  $\int_\Sigma w\,\delta(k_i\cdot P)  \e^{ik_i\cdot x}$ in the vector representation where $w$ is some worldsheet matter that depends on the polarization data. Using the incidence relation, this can be obtained from \begin{equation}
\int d^2u_i\, d^2v_i\, \delta\big(\langle v_i \varepsilon_i \rangle -1 \big)\;\delta^4\Big(\langle u_{i}\lambda_A(\sigma_i)\rangle -\langle v_{i}\kappa_{iA}\rangle\Big)w\,\e^{ i\langle\mu^{A}u_i\rangle \epsilon_{iA}}\, ,\label{plane-wave}
\end{equation}
in the twistor representation after integrating out $(u,v)$ against the delta functions. This vertex operator has cohomology degree one since one delta-function remains after integrating out the $(u_i,v_i)$.\footnote{The delta functions are most naturally expressed as Dolbeault $(0,1)$-forms by $\delta(z)=\bar\p \frac1{2\pi iz}$.} Moreover, by the discussion in \cref{sec:polSE}, this remaining delta-function is a multiple of $\delta(k_i\cdot P)$, and the exponent  reduces to 
\begin{equation}
\mu^{Aa}u_{ia}\epsilon_{iA} =x^{AB}\lambda_B^au_{a} \epsilon_{iA}=x^{AB}\kappa_{iaB}v^a_i\epsilon_{iA}=k_i\cdot x \, ,
\end{equation}
on the support of the delta functions. The first equality makes use of the incidence relations \eqref{A6dinc}, and the last equality follows from the normalization  $\epsilon_{[A}\kappa_{B]a}v^a=k_{AB}$. 

This general form for the vertex operator is sufficient to derive the polarized scattering equations from the ambitwistor string path integral. Consider the correlation function of $n$ vertex operators, indexed by $i=1,\ldots, n$. The $(\lambda,\mu)$ path integral in this correlator can be performed by taking the exponentials into the action to yield sources in the equation for $\lambda$:
\begin{equation}
\dbar \lambda_{aA}=\sum_i \epsilon_{iA} u_{ia}\bar \delta(\sigma-\sigma_i)\, .
\end{equation}
Since $\lambda$ is a worldsheet spinor, this has the unique solution 
\begin{equation}\label{equ:lambda}
\lambda_A^a(\sigma)=\sum_{i=1}^n \frac{u_{i}^a\, \epsilon_{iA}}{\sigma-\sigma_i}\, ,
\end{equation}  
justifying our Ansatz in \eqref{lambda-def}. The delta functions then impose the polarized scattering equations \eqref{polscatt} irrespective of the worldsheet matter $W$ whose correlators give rise  to  the integrands $\mathcal{I}_n$.
%;
%\begin{equation}
%\sum_{j=1}^n\frac{\langle u_{i}u_{j}\rangle\epsilon_{jA}}{\sigma_{ij}}  = \langle v_{i}\kappa_{iA}\rangle\, ,
%\end{equation}
%which are nonsingular  because $\epsilon^{ab}$ is skew. The vertex operators of the ambitwistor string are thus well defined, and supported on the polarized scattering equations.

\subsection{Supersymmetric ambitwistor strings}
Extend space-time to superspace to include fermionic coordinates $\theta^{AI}$ with $I=1,\dots,2{N}$. Supertwistor coordinates are then $\cZ=(\lambda_A,\mu^A,\eta^I)$, \cite{Chern:2009nt}.  For our supertwistor description of super-ambitwistor space we use a pair of supertwistors $\cZ^a$ with  incidence relations
\begin{equation}
\mu^{Aa}=x^{AB}\lambda_B^a + \theta^{IA}\theta_I^B\lambda_B^a\, , \qquad \eta^{Ia}=\theta^{IA}\lambda^a_{A}\, ,
\end{equation}
subject to the constraints
\begin{equation}
0=\cZ^a\cdot \cZ^b:=\mu^{Aa}\lambda^{b}_A+\mu^{Ab}\lambda^{a}_A +\Omega_{IJ}\,\eta^{Ia}\eta^{Jb}\, . 
\end{equation}
%Ambidextrous supersymmetry can be introduced analogously, with $\tilde \eta^{\dot I}_{\dot a}=\tilde \theta^{\dot I}_A\lambda_{\dot a}^A$. Here, we focus on the chiral version, the extension to ambidextrous supersymmetry is straightforward. 

In the superambitwistor string, $\eta^I_a$ become spinors on the world sheet along with the other components of $\cZ^a$ leading to  worldsheet action
%, and the full supersymmetric action is given by $S_{\scalebox{0.7}{s$\mathbb{A}$}}^{\,\scalebox{0.6}{$(N,\tilde N)$}}=S_{\scalebox{0.7}{$\mathbb{A}$}}+S_{\eta}^{\scalebox{0.6}{$(N)$}}+S_{\tilde\eta}^{\scalebox{0.6}{$(\tilde N)$}}$, with
\begin{equation}
 S^{\scalebox{0.6}{$(N)$}}=\int_\Sigma \epsilon_{ab}\cZ^a\bar\p \cZ^b + A_{ab}\cZ^a\cdot \cZ^b
%\Omega_{IJ}\left(\epsilon_{ab}\;\eta^{Ia}\bar\p\eta^{Ja}+ A_{ab}\;\eta^{Ia}\eta^{Jb}\right)\,.
\end{equation}

\smallskip

The bosonic vertex operators \eqref{plane-wave} extend to ones for  supermomentum eigenstates via
\begin{equation}
\int d^2u_i\, d^2v_i\, \delta\big(\langle v_i \varepsilon_i \rangle -1 \big)\;\delta^4\Big(\langle u_{i}\lambda_A(\sigma_i)\rangle -\langle v_{i}\kappa_{iA}\rangle\Big)\,w\,\e^{ iu_{i}^a\left(\mu^{A}_a \epsilon_{iA}+\eta^{I}_a q_{iI}\right)}\, .\label{susy-estate}
\end{equation}
To see this, first observe that  the defining relations \eqref{eq:def_susy-gen} for supermomenta $(q_I,\tilde q_{\dot I})$ are solved by $ \exp (ik\cdot x+q_I\theta^{IA}\kappa_A^av_a +\Omega_{IJ}\theta^{IA}\theta^{JB}k_{AB})$.
\iffalse
\begin{equation}
 \exp (ik\cdot x+q_I\theta^{IA}\kappa_A^av_a +\Omega_{IJ}\theta^{IA}\theta^{JB}k_{AB})\, .
\end{equation}\fi
Then the incidence relations and polarized scattering equations identify the exponent with  that in \eqref{susy-estate}.

For an $n$-particle amplitude, the $\eta$ path-integral can  be computed as before by taking the exponentials into the actions giving  sources for the $\eta$ equation of motion $\bar\p \eta^{Ia} =\frac{1}{2}\sum_iu_i^a q_{i}^I \bar\delta(\sigma-\sigma_i)$. This is solved by
\begin{equation}\label{eq:eta}
\eta^{Ia}(\sigma)=\frac{1}{2}\sum_i \frac{u_i^aq^I_i}{\sigma-\sigma_i}\,.
\end{equation}
The remaining contribution to the amplitude is via the exponential factor in \eqref{susy-estate} which reduces, after substituting the solution \eqref{eq:eta}, to give
\begin{equation}
\exp\left(F_{\scalebox{0.65}{$N$}}\right)=\exp \left( \sum_{i<j}   \frac{ \langle u_i u_j\rangle q_{iI}q_j^I}{\sigma_{ij}}\right)\, .
\end{equation}

\section{\texorpdfstring{$\mathcal{N}=4$}{N=4} super Yang-Mills  on the Coulomb branch} \label{sec:CB}
The 6d $(1,1)$ SYM amplitudes \eqref{eq:int_sYM} can be used directly to obtain amplitudes for 4d $\mathcal{N}=4$ SYM  on the Coulomb branch  \cite{Bern:2010qa,Cachazo:2018hqa}. %Huang:2011um
  The  4d massive kinematic data is $(\lambda_{\alpha a},\tilde \lambda_{\dot \alpha a})$ where  $a=0,1$ is the 4d massive little group index \cite{Arkani-Hamed:2017jhn}, and
\begin{equation}
 \lambda_{\alpha a}\tilde\lambda_{\dot\alpha b}\epsilon^{ab}=k_{\alpha\dot\alpha}\,,\qquad \lambda_{\alpha a}\lambda_{\beta b}\epsilon^{ab} = M\epsilon_{\alpha\beta}\,,\qquad \tilde\lambda_{\dot\alpha a}\tilde\lambda_{\dot\beta b}\epsilon^{ab} = \tilde{M}\epsilon_{\dot\alpha\dot\beta}\,.
\end{equation}
On reduction to 4d, the spinor index can be broken up into two-component 4d spinor indices as $A=(\alpha,\dot\alpha)$ and the two $SU(2)$ factors of the 6d little group can both be identified with the 4d massive one, so  $a=\dot a$. This 
embeds into the 6d spinor-helicity formalism, reinterpreting some of the spinor components of the momenta as masses, as
\begin{equation}\label{eq:red_massive_4d}
 \kappa_{aA}=\begin{pmatrix}\lambda_{\alpha a}\\ 
  \tilde\lambda^{\dot\alpha}_a\end{pmatrix}\,,\qquad \kappa_{\dot a}^A=\kappa^A_a=\begin{pmatrix}\lambda_{a}^\alpha \\ \tilde\lambda_{\dot\alpha a}\end{pmatrix}\,.
\end{equation}
with $M\tilde M=m^2$.  The reduction to massless kinematics follows by setting $\lambda_{\alpha,0} = \tilde\lambda_{\dot\alpha,1}=0$.

For 6d $(1,1)$-SYM, masses are introduced  by giving some of the scalars a non-zero vev $\phi_0$
%, \footnote{Explicitly, we take $\langle(\phi_{I \dot I})^{J_1}_{J_2}\rangle =v \,(\delta_I^0\delta_{\dot I}^{\dot 0}+\delta_I^1\delta_{\dot I}^{\dot 1})\,\delta^{J_1}_{J_2} $,  where  $J_{1,2}$ denote $\mathrm{U}(M)$ indices.} 
that spontaneously breaks the gauge group from U$(N+M)$ to U$(N)\times \mathrm{U}(M)$. This gives a mass $m=g_{\mathrm{YM}}\, \phi_0$ to the off-diagonal gauge bosons, now bi-fundamentals under both gauge groups.

%Define $\lambda_{i\alpha} = v_{ia}\lambda_{i\alpha}^a$ and $\epsilon_{i\alpha} = \epsilon_{ia}\lambda_{i\alpha}^a$, and similarly for $\tilde\lambda_{i}^{\dot\alpha}$, $\tilde\epsilon_{i}^{\dot\alpha}$ and 
Denote now little group contractions by round brackets, e.g. $(u_iu_j):= u_{ia}u_j^a$ and reserve now the angle and square bracket for the usual 4d spinor-helicity contractions. In this notation, the polarized  massive 4d scattering equations are given by 
\begin{subequations}\label{eq:SE_CB}
\begin{align}
% \mathcal{E}_{i\alpha} = 
 \sum_j\frac{( u_i u_j)}{\sigma_{ij}}\epsilon_{j\alpha} = v_{ia}\lambda^a_{i\alpha} \,,\qquad
% \tilde{\mathcal{E}}_i{}^{\dot\alpha} = 
\sum_j\frac{( u_i u_j)}{\sigma_{ij}}\tilde \epsilon_{j}{}^{\dot\alpha} = v_{ia}\tilde\lambda_{i}^{\dot\alpha a}\,.
\end{align}
\end{subequations}
%where $\epsilon_{i\alpha}=(\epsilon_i\,\lambda_{i\alpha})$ and $\tilde\epsilon_{i}^{\dot \alpha}=(\tilde \epsilon_i\,\tilde \lambda_{i}^{\dot \alpha})$.
Both the chiral and the anti-chiral 6d polarised scattering equations descend to the same massive 4d scattering equations, due to the identification \eqref{eq:red_massive_4d} (this is already a feature in 5d). 

The matrix $H$ used to define the SYM integrand also reduces straightforwardly,  becoming a symmetric matrix with entries given by
\begin{equation}
 H^{\mathrm{CB}}_{ij} = \frac{\left\langle \epsilon_i\epsilon_j  \right\rangle+ \left[ \tilde\epsilon_i\tilde\epsilon_j  \right]}{\sigma_{ij}}\,,
\end{equation}
with its reduced determinant defined as before as $\det{}'H^{\mathrm{CB}} =(u_{i_1}u_{i_2})^{-1}(u_{j_1}u_{j_2})^{-1}\det H^{\mathrm{CB}}{}^{[i_1i_2]}_{[j_1j_2]}$.
\iffalse
\begin{equation}
  \det{}'H^{\mathrm{CB}} %= \frac{\det H^{\mathrm{CB}}{}^{[ij]}_{[ij]}}{(u_iu_j)^2} 
  =  \frac{\det H^{\mathrm{CB}}{}^{[i_1i_2]}_{[j_1j_2]}}{(u_{i_1}u_{i_2})(u_{j_1}u_{j_2})}  \,.
\end{equation}\fi
The full scattering amplitude of $\mathcal{N}=4$ SYM on the Coulomb branch is then given by
\begin{equation}
 \mathcal{A}_n^{\mathrm{CB}} = \int \rd\mu_n^{\mathrm{CB}}\; \mathrm{PT}(\alpha) \det{}'H^{\mathrm{CB}}\;e^{F_1+\tilde{F}_1}\,,
\end{equation}
where the measure $\rd\mu_n^{\mathrm{CB}}$ on the Coulomb branch is given by \eqref{eq:measure} but with the delta-functions now imposing  the massive 4d polarized scattering equations \eqref{eq:SE_CB}.

\section{Discussion}\label{sec:discussion}
The Coulomb branch amplitudes will also be subject to some of the awkwardnesses of the massive CHY amplitudes as proposed by Naculich \cite{Naculich:2014naa}.    
The further reduction of the Coulomb branch to the massless sector leads to the refined scattering equations of \cite{Geyer:2014fka} which also incorporate polarization data, although this is now just a scaling as the massless little group in 4d is now $\C^*\times \C^*$. The refinement by MHV degree is very much a 4d phenomena.  The reduced determinant $\det'H$ then reduces on the one hand to the reduced determinant of the Hodge matrices for gravity in  \cite{Geyer:2014fka} and on the other to the Jacobian between the refined/polarized scattering equations and the CHY measure (and hence the reduced polarized scattering equations by \eqref{eq:measure}). Thus our formulae reduce to the correct formulae in 4d.  This is mapped in  \cite{Geyer:2016nsh} to the
well-known twistor-string super Yang-Mills amplitudes \cite{Witten:2003nn,Berkovits:2004hg, Roiban:2004yf}; it is these latter formmulae that relate most closely to the  4d reduction of \cite{Cachazo:2018hqa}.

One major distinction between our work and that of \cite{Cachazo:2018hqa} is that there is now no artificial  distinction between even and odd particles in the polarized scattering equations. The only distinction between even and odd particle numbers in our work is physical, as seen in the integrands for M5 and D5 theory based on Pfaffians $\pf{}'A$ of skew $n\times n$ matrices that  vanish when $n$ is odd. 
 Our choice of supersymmetry has the advantage that it does not rely on the breaking of Lorentz or R-symmetry.  It is however  dynamic changing from particle to particle, and so some fermionic Fourier transform is required to relate it to more conventional global representations.  This will often lead to fermionic delta functions that provide fermionic analogues of the polarized scattering equations for the $\eta$'s. 

%There are no associated scattering equations\footnote{One can transform to the supermomenta representation $(q_{\mathrm{I}}^a,\tilde q^{\dot a}_{\dot{\mathrm{I}}})$ that manifests little group invariance by an appropriate Grassmann Fourier transform.  This leads to fermionic delta functions that impose $\langle u_i \eta^{\mathrm{I}}(\sigma_i)\rangle =\langle v_i q_i^{\mathrm{I}}\rangle$ with $\mathrm{I}=1,\dots,\mathcal{N}$.} due to the form of the supermomentum eigenstate, so 

\smallskip

There are many directions for future investigation. In light of the `bare' ambitwistor string model presented here, a natural question is whether there are worldsheet matter systems describing amplitudes in the theories presented here. For ambidextrous theories such as super Yang-Mills and supergravity, this seems to require both chiral and antichiral supertwistors. As indicated above, these are alternative coordinatizations of the same space,  requiring an implementation of the associated constraints in the worldsheet model. Provided this is resolved, there seems to be a natural candidate for the matter system giving rise to the $\det{}'H$ factors, constructed from fermionic fields $(\rho_A, \rho^A)\in K^{1/2}\otimes(\mathbb{S}_A,\mathbb{S}^A)$;
\begin{equation}
S_{\rho}=\int_{\Sigma} \rho_A\bar\p \rho^A + \tilde B^{\dot a} \lambda^A_{\dot a} \rho_A + B_a \lambda^a_A\rho^A\,,
\end{equation}
where $B$ and $\tilde B$ are fermionic $(0,1)$-form gauge fields on the worldsheet. For this system there are simple natural choices for the worldsheet matter $W$ in the vertex operators that give rise to $\det'H$.  We leave an exploration of this  along with analogues in higher dimensions for future work \cite{wip}.

A different different for future research concerns what other theories can be constructed using the polarized scattering equations. One immediate observation is that several new  formulae can be constructed  by combining the half-integrands discussed in \cref{sec:integrands}:
\begin{center}
\begin{tabular}{l|llll}
 & PT & $\det{}'A$ & $\det{}'H \;e^{F_1+\tilde{F}_1}$ & $\displaystyle \frac{\pf{}'A\,\det X}{\det^{2} U}\; e^{F_2}$\vspace{3pt}\\\hline
 PT & Bi-adjoint scalar &  NLSM & $\mathcal{N}=(1,1)$ SYM & $\mathcal{N}=(2,0)$\\
 $\det{}'A$ &&Galileon& $\mathcal{N}=(1,1)$ D5 & $\mathcal{N}=(2,0)$ M5 \\
 $\det{}'H \;e^{F_1+\tilde{F}_1}$ &&&$\mathcal{N}=(2,2)$ sugra& $\mathcal{N}=(3,1)$\\
$\displaystyle \frac{\pf{}'A\,\det X}{\det^{2} U}\; e^{F_2}$ &&&& $\mathcal{N}=(4,0)$
\end{tabular}
\end{center}
In particular, combining the `M5' half-integrand with a Parke-Taylor factor leads to a formula with a non-abelian current algebra and $\mathcal{N}=(2,0)$ supersymmetry. As observed in \cite{Cachazo:2018hqa} for a similar construction, this cannot describe amplitudes in the $\mathcal{N}=(2,0)$ theory arising from coincident M5-branes due to explicit no-go theorems, and the absence of a perturbative parameter. The formulae are however well-defined and manifestly chiral and supersymmetric, so it is worth studying these expressions in their own right. %, including the related $\mathcal{N}=(4,0)$ and $\mathcal{N}=(3,1)$  `supergravity' formulae. %In particular, it would be interesting to compare to the results in \cite{Cachazo:2018hqa}, suggesting a peculiar non-local behaviour under factorization.

\medskip

\emph{Acknowledgements:} We would like to thank the authors of \cite{Cachazo:2018hqa} for freely discussing their work when it was in its early stages and sharing drafts, and for valuable discussions. YG gratefully acknowledges support from the National Science Foundation Grant PHY-1606531 and the Association of Members of the Institute for Advanced Study (AMIAS). LJM  is grateful to the EPSRC for support under grant EP/M018911/1.

\iffalse
\begin{subequations}\label{eq:integrands_proposal}
 \begin{align}
  & \mathcal{N}=(2,0): && \mathrm{PT}(\alpha)\;\frac{\pf{}'A\,\det X}{\det^{2} U}\; e^{F_2}\\
  & \mathcal{N}=(3,1): && \det{}'A\;\frac{\pf{}'A\,\det X}{\det^{2} U}\; e^{F_3+\tilde F_1}\\
  & \mathcal{N}=(4,0): && \left(\frac{\pf{}'A\,\det X}{\det^{2} U}\right)^2\; e^{F_4}
 \end{align}
\end{subequations}\fi

\bibliography{twistor-bib}
\bibliographystyle{JHEP}

\end{document}